\documentclass[11pt,a4paper]{article}

\usepackage[T1]{fontenc}
\usepackage[utf8]{inputenc}
\usepackage{lmodern}
\usepackage{microtype}
\usepackage[a4paper,margin=2.25cm]{geometry}
\usepackage{amsmath,amssymb,bm}
\usepackage{graphicx}
\usepackage{float}
\usepackage{booktabs}
\usepackage[dvipsnames]{xcolor}
\usepackage[numbers,sort&compress]{natbib}
\usepackage{hyperref}
\usepackage{doi}
\hypersetup{
  colorlinks=true,
  linkcolor=MidnightBlue,
  citecolor=MidnightBlue,
  urlcolor=MidnightBlue,
  pdftitle={Displaced e+e- Signatures from an $X_{17}$-Mediated Inelastic Dark-Sector Cascade},
  pdfauthor={Raoul Serao and Antonio Capolupo}
}

\newcommand{\eps}{\varepsilon}
\newcommand{\mX}{m_X}
\newcommand{\chiH}{\chi_{\mathrm H}}
\newcommand{\chiL}{\chi_{\mathrm L}}
\newcommand{\mH}{m_{\mathrm H}}
\newcommand{\mL}{m_{\mathrm L}}
\newcommand{\ctau}{c\tau}

\newcommand{\DeltaM}{\Delta}

\title{Displaced $e^+e^-$ Signatures from an $X_{17}$-Mediated Inelastic Dark-Sector Cascade}

\author{
Raoul Serao$^{1,*}$ and Antonio Capolupo$^{2,+}$\\[0.65em]
\small $^1$Department of Physics, Faculty of Science, University of Zagreb,\\[-0.2em]
\small Bijeni\v{c}ka cesta 32, 10000 Zagreb, Croatia\\
\small $^2$Dipartimento di Fisica E.~R. Caianiello, Universit\`a di Salerno,\\[-0.2em]
\small and INFN -- Gruppo Collegato di Salerno, Via Giovanni Paolo II 132, 84084 Fisciano (SA), Italy\\[0.45em]
\small $^*$Correspondence: rserao@unisa.it, $+$capolupo@sa.infn.it
}

\date{}

\begin{document}
\maketitle

\begin{abstract}

We investigate displaced electron–positron signatures from an inelastic dark sector mediated by a light vector boson associated with the $X_{17}$ anomaly. A heavy neutral fermion decays into a lighter stable state and the mediator, whose subsequent decay produces a single visible vertex after two stages of neutral propagation. We calculate the decay widths, displacement distribution, lepton kinematics, and production rate in electron–positron collisions, including vector polarization. At Belle II, stronger dark-sector couplings increase production but require finely adjusted masses to preserve a macroscopic decay length. We quantify off-shell contributions and assess the validity of the two-body approximation. We show that the electron–positron pair cannot satisfy simultaneously the momentum and opening-angle cuts of an existing search proposal for the configuration studied. Less restrictive selections allow nonzero kinematic acceptance. Our results clarify the experimental requirements for searching for these signatures at Belle II.

\end{abstract}


\section{Introduction}
\label{sec:introduction}
An excess of electron--positron pairs in an excited $^8$Be transition motivated the proposal of a new boson with a mass near $17$ MeV~\cite{Krasznahorkay2016}. Similar anomalies were subsequently reported in $^4$He and $^{12}$C transitions~\cite{Krasznahorkay2021,Krasznahorkay2022,Krasznahorky2024}. Several theoretical interpretations have been proposed~\cite{Kozaczuk2017,Feng2020,Viviani2024,Fieg2026,Batra2026,CapolupoX2024,X4}, including light vector bosons with approximately protophobic couplings~\cite{Feng2016,Feng2017,Barducci2023}. The existence of the $X_{17}$ boson remains unconfirmed~\cite{Alves2023,Gustavino2024,X3}. MEG~II found no significant signal~\cite{MEGII2025}, while PADME investigated resonant production near the proposed mass~\cite{PADME2025}. A recent combination of the available results gives $m_X=16.88\pm0.05$ MeV under specific assumptions~\cite{Combined2026}. Other experiments constrain its possible interactions with Standard Model particles~\cite{KLOE2015,NA642018,NA642020,FASER2024,E141update2026,NA482015,SINDRUM1992,Lees2014DarkPhoton,Riordan1987,Bjorken1988}.

A light vector boson could also mediate interactions between Standard Model particles and an inelastic dark sector~\cite{Holdom1986,Batell2009,Foguel2025}. Such models contain dark states with different masses, connected through off-diagonal interactions~\cite{TuckerSmithWeiner2001,Izaguirre2015,CarrilloGonzalez2022,Filimonova2022,Izaguirre2017}. The heavier state can decay into the lighter one through the emission of a vector mediator. If the mediator is produced off shell, the decay generates an electron--positron pair at the heavy-state decay point. If it is produced on shell, it can travel before decaying. The visible vertex is then displaced from the heavy-state decay point. Both regimes have been investigated in previous studies~\cite{Giudice2018,BerlinKling2019,Duerr2020,BatellSBN2021,Duerr2021}.

We consider two neutral fermions, a heavy state $\chiH$ of mass $m_H$ and a lighter stable state $\chiL$ of mass $m_L$. They interact through a vector boson $X$ of mass $m_X$ in the region associated with the $X_{17}$ anomaly. When the mass difference $m_H-m_L$ exceeds $m_X$, the heavy state can decay through $\chiH\to\chiL X$. The mediator subsequently decays into an electron--positron pair. The first decay produces no charged particles. Only the second decay generates a visible vertex. Its position therefore depends on the propagation of both $\chiH$ and $X$.

We calculate the decay widths, the distribution of the visible vertex, and the lepton kinematics, including vector polarization. We also determine the production rate in electron--positron collisions. The process produces two invisible dark fermions and a displaced electron--positron pair.

Our analysis reveals a tension between production and decay length. Increasing the dark-sector coupling enhances particle production but reduces the heavy-state lifetime at fixed masses. Maintaining a macroscopic displacement requires the mass difference to approach the mediator mass with increasing precision. We define the mass excess as $\delta\equiv m_H-m_L-m_X$ and show that it can reach the electronvolt scale. This relation is not protected by a symmetry in the effective model. We also calculate the off-shell contribution to the decay width and establish where the two-body approximation remains reliable.

We apply these results to Belle~II. For the mass configuration studied, we show analytically that the electron--positron pair cannot simultaneously satisfy the momentum and opening-angle requirements of the displaced-pair selection proposed in Ref.~\cite{Duerr2020}. This incompatibility persists even without the invariant-mass cut. We then examine alternative selections and identify conditions that retain a nonzero fraction of the signal. Our results establish the kinematic requirements for dedicated searches for displaced low-mass electron--positron pairs. A complete assessment of experimental sensitivity requires detector-level studies~\cite{Lee2019,Alimena2020}.

The paper is organized as follows. In Sec.~\ref{sec:theory}, we introduce the minimal inelastic dark-sector framework and derive the heavy-state and mediator decay properties, the two-stage displacement distribution, and the relevant kinematic and polarization observables. In Sec.~\ref{sec:phenomenology}, we connect these results to the physical collider process, compute the associated production rate, and investigate the phenomenological feasibility of the displaced $e^+e^-$ signature at Belle~II. Finally, Sec.~\ref{sec:conclusions} summarizes our main results and conclusions.

\section{Minimal inelastic dark sector}
\label{sec:theory}
We consider two neutral Dirac fermions, $\chiL$ and $\chiH$, with masses $\mL<\mH$ and mass difference $\DeltaM\equiv\mH-\mL$. We assume the lighter state is stable and interacts too weakly with detector material to be reconstructed. The fermions interact through a vector boson $X$ of mass $\mX$:
\begin{equation}
\mathcal L_{\rm dark}
\supset
 g_D X_\mu\bar\chiL\gamma^\mu\chiH+\mathrm{h.c.}
\label{eq:darklag}
\end{equation}
Here $\mathcal L_{\rm dark}$ is the dark-sector interaction Lagrangian density. The parameter $g_D$ is real and dimensionless. The field $X_\mu$ describes the vector mediator, $\gamma^\mu$ are Dirac matrices, and $\mu$ is a spacetime index. Bars denote Dirac adjoints; $\mathrm{h.c.}$ denotes the Hermitian conjugate. Such interactions occur in inelastic dark-sector models~\cite{TuckerSmithWeiner2001,Izaguirre2015,CarrilloGonzalez2022,Filimonova2022}. We take the mediator coupling to electrons to be
$\mathcal L_{e}
=
 e\eps_e X_\mu\bar e\gamma^\mu e,$
where $e$ is the electromagnetic coupling and $\eps_e$ is its dimensionless scaling factor for $X$. We assume that $X$ decays only to $e^+e^-$ and that $\chiH$ has no additional decay interactions. We include the off-shell channel $\chiH\to\chiL e^+e^-$ when computing the total heavy-state width.

The decay $\chiH\to\chiL X$ is allowed for $\DeltaM>\mX$. We define the mass excess $\delta\equiv\DeltaM-\mX>0$. The spin-averaged squared amplitude is
\begin{equation}
\overline{|\mathcal M|^2}
=
 g_D^2
\frac{\left[(\mH-\mL)^2-\mX^2\right]
\left[(\mH+\mL)^2+2\mX^2\right]}{\mX^2}.
\label{eq:amplitudeH}
\end{equation}
We use natural units ($\hbar=c=1$) for amplitudes and decay widths. The corresponding two-body partial width is~\cite{PDG2024}
\begin{equation}
\Gamma(\chiH\to\chiL X)
=
\frac{g_D^2}{16\pi\mH^3\mX^2}
\lambda^{1/2}(\mH^2,\mL^2,\mX^2)
\left[(\mH-\mL)^2-\mX^2\right]
\left[(\mH+\mL)^2+2\mX^2\right],
\label{eq:gammaH}
\end{equation}
where $\lambda(x,y,z)=x^2+y^2+z^2-2xy-2xz-2yz$ is the K\"all\'en function. We denote this partial width by $\Gamma_2$. The corresponding estimate of the proper decay length is
$\ctau_{\chiH}^{(2)}=\frac{\hbar c}{\Gamma_2},$
where $\hbar$ is the reduced Planck constant, $c$ is the speed of light, and $\tau_{\chiH}$ is the proper lifetime of $\chiH$. This estimate equals the physical length $\ctau_{\chiH}$ only when the on-shell contribution dominates the total width. We test this condition in Sec.~\ref{sec:phenomenology}.

\begin{figure}[H]
\centering
\includegraphics[width=0.66\textwidth]{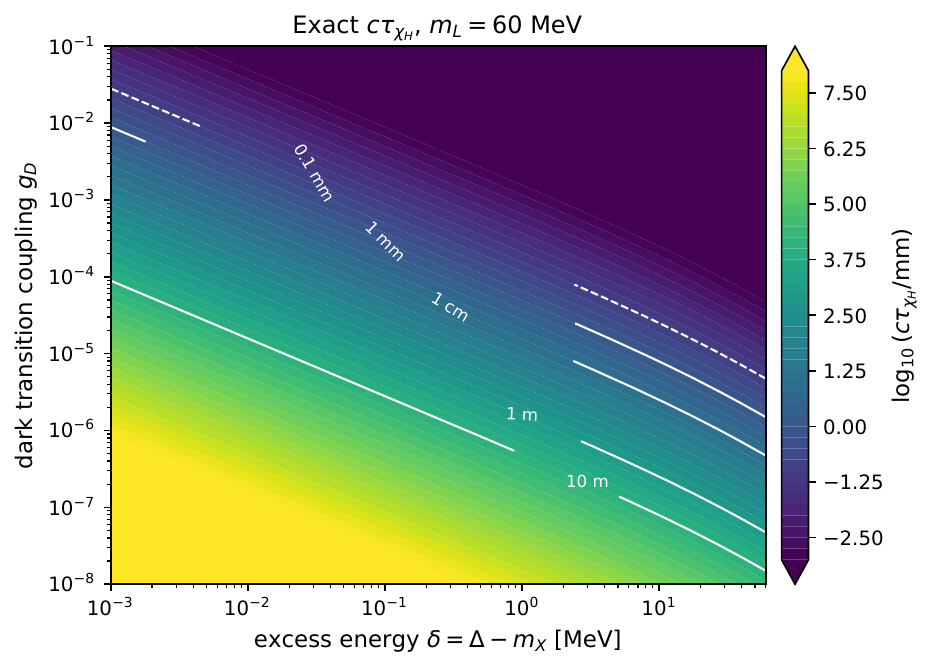}
\caption{Two-body proper decay length $\ctau_{\chiH}^{(2)}$ in the $(\delta,g_D)$ plane for $\mL=60$ MeV and $\mX=16.88$ MeV. White curves mark lengths from $0.1$ mm to $10$ m.}
\label{fig:chiHmap}
\end{figure}

For $\delta\ll\mX,\mL$, the width scales as $\delta^{3/2}$:
$\Gamma(\chiH\to\chiL X)
\simeq
 g_D^2\,A(\mL,\mX)\,\delta^{3/2},$
where $A(\mL,\mX)$ is a mass-dependent coefficient:
\begin{equation}
A(\mL,\mX)
=
\frac{\left[(2\mL+\mX)^2+2\mX^2\right]
\sqrt{8\mL\mX(\mL+\mX)}}
{8\pi(\mL+\mX)^3\mX}.
\end{equation}
Thus $\ctau_{\chiH}^{(2)}\propto g_D^{-2}\delta^{-3/2}$. A long-lived heavy state requires either a small coupling or a small mass excess. Figure~\ref{fig:chiHmap} shows this dependence for $\mL=60$ MeV and $\mX=16.88$ MeV. At fixed two-body decay length, increasing $g_D$ requires reducing $\delta$. For example, $g_D=10^{-2}$ gives a two-body proper length of $1$ cm at $\delta\simeq1.84\times10^{-4}$ MeV.

The small excess is $\delta$, not the full splitting $\DeltaM$. For the principal Belle~II parameter choice studied below, $\delta=8.54$ eV. Setting $\delta=0$ does not restore a symmetry of the effective Lagrangian. The mass relation therefore has no established symmetry protection~\cite{tHooft1980}. We treat the eV-scale excess as a tuned input; its radiative stability would require a specified ultraviolet completion. We adopt $\mX=16.88$ MeV, with a quoted uncertainty of $0.05$ MeV from a conditional combination of measurements~\cite{Combined2026}. Varying $\mX$ within this uncertainty can close the on-shell channel for sufficiently small $\delta$.

The mediator decays into an electron--positron pair with partial width~\cite{Pospelov2009,Bjorken2009}
\begin{equation}
\Gamma(X\to e^+e^-)
=
\frac{\alpha\eps_e^2\mX}{3}
\left(1+\frac{2m_e^2}{\mX^2}\right)
\sqrt{1-\frac{4m_e^2}{\mX^2}},
\label{eq:gammaX}
\end{equation}
where $\alpha=e^2/(4\pi)$ is the fine-structure constant and $m_e$ is the electron mass. Our assumption of a single decay mode gives the total width $\Gamma_X=\Gamma(X\to e^+e^-)$ and the proper decay length $\ctau_X=\hbar c/\Gamma_X$, where $\tau_X$ is the proper lifetime of $X$. For the adopted mediator mass,
$\ctau_X\simeq
0.4806~\mathrm{mm}
\left(\frac{10^{-4}}{|\eps_e|}\right)^2.$
An updated E141 analysis~\cite{E141update2026} reports no exclusion near this mass for a vector mediator with negligible neutrino coupling and unit branching fraction $\mathrm{Br}(X\to e^+e^-)=1$. Its comparison with NA64 and Orsay constraints~\cite{NA642020,Davier1989} gives the illustrative range $6.5\times10^{-5}\lesssim|\eps_e|\lesssim1.1\times10^{-4}$. We use this range conditionally, not as a generally allowed region~\cite{Ilten2018,Bauer2018,HostertPospelov2023,X2}. It corresponds to $0.397~\mathrm{mm}\lesssim\ctau_X\lesssim1.138~\mathrm{mm}$.

A separate condition concerns the on-shell approximation. The mass excess must exceed the mediator width, $\delta\gg\Gamma_X$. This condition alone does not guarantee that the two-body channel dominates the inclusive decay $\chiH\to\chiL e^+e^-$~\cite{Kauer2007,Berdine2007,Kauer2008}. We quantify the off-shell correction in Sec.~\ref{sec:phenomenology}.

We now calculate the position of the visible vertex. Unless stated otherwise, all distances and directions refer to the laboratory frame. Let $\mathbf r_0$ be the production point of $\chiH$. The heavy state travels a distance $\ell_H$ along $\hat{\mathbf n}_H$. The mediator then travels $\ell_X$ along $\hat{\mathbf n}_X$. Both $\hat{\mathbf n}_H$ and $\hat{\mathbf n}_X$ are unit vectors. The reconstructed $e^+e^-$ vertex lies at
$\mathbf r_{ee}
=
\mathbf r_0
+\ell_H\hat{\mathbf n}_H
+\ell_X\hat{\mathbf n}_X,
\qquad
R_{ee}
=
\left|\mathbf r_{ee}-\mathbf r_0\right|.$
Here $R_{ee}$ is the three-dimensional displacement from production to the visible vertex. The first decay is invisible. The measured displacement therefore contains both neutral flight paths.

For a fixed heavy-state boost, the mean laboratory flight distances are
$\overline\ell_H
=
\beta_H\gamma_H\ctau_{\chiH},
\qquad
\overline\ell_X(\theta_X)
=
\frac{p_X^{\rm lab}(\theta_X)}{\mX}\ctau_X.$
Here $\beta_H$ is the heavy-state speed in units of $c$, $\gamma_H=(1-\beta_H^2)^{-1/2}$, and $\ctau_{\chiH}$ and $\ctau_X$ are the physical proper decay lengths. The mediator emission angles $(\theta_X,\phi_X)$ are defined in the $\chiH$ rest frame. The polar angle $\theta_X$ is measured from the heavy-state laboratory direction; $\phi_X$ is the azimuth around that axis. Its laboratory three-momentum magnitude is $p_X^{\rm lab}(\theta_X)=|\mathbf p_X^{\rm lab}(\theta_X,\phi_X)|$; rotational symmetry makes it independent of $\phi_X$.

For an unpolarized $\chiH$, the emission is isotropic in its rest frame. At fixed angles, $\ell_H$ and $\ell_X$ are independent exponential random variables. Their joint probability density is
\begin{equation}
\frac{d^4P}
{d\ell_H\,d\ell_X\,d\cos\theta_X\,d\phi_X}
=
\frac{
\exp\!\left(-\ell_H/\overline\ell_H\right)
\exp\!\left[-\ell_X/\overline\ell_X(\theta_X)\right]
}{
4\pi\,\overline\ell_H\,\overline\ell_X(\theta_X)
}.
\label{eq:displacementkernel}
\end{equation}
Here $P$ is the decay probability, with $\ell_H,\ell_X\geq0$, $-1\leq\cos\theta_X\leq1$, and $0\leq\phi_X<2\pi$.

\begin{figure}[H]
\centering
\includegraphics[width=0.98\textwidth]{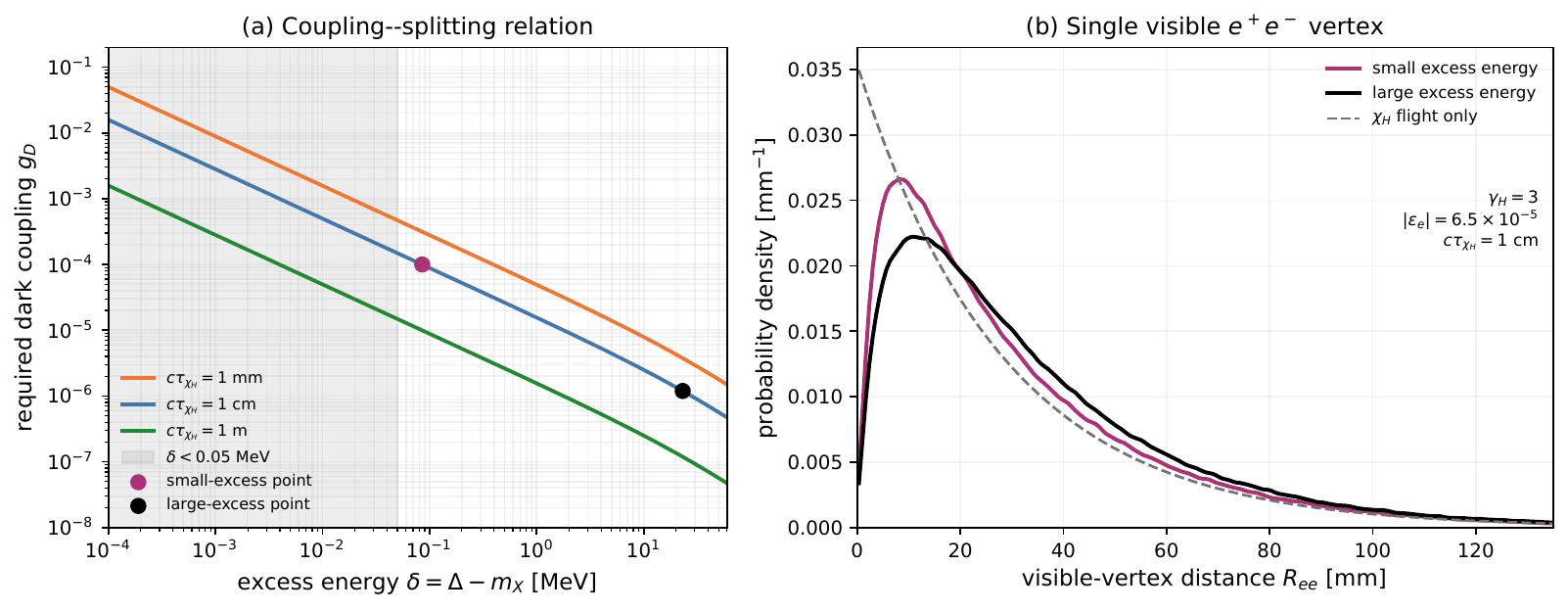}
\caption{Left: coupling required for fixed two-body heavy-state decay lengths as a function of $\delta$. The gray region marks $\delta<0.05$ MeV, the quoted mediator-mass uncertainty scale~\cite{Combined2026}. Both marked points have $\ctau_{\chiH}^{(2)}=1$ cm. Right: visible-vertex displacement for a fixed boost $\gamma_H=3$ and $|\eps_e|=6.5\times10^{-5}$. The distributions use $7.5\times10^5$ decays generated from Eq.~\eqref{eq:displacementkernel}, without detector effects.}
\label{fig:observable_displacement}
\end{figure}

A small mass excess suppresses the mediator recoil. A superscript $*$ denotes the $\chiH$ rest frame, and $\mathbf p_X^*$ is the mediator three-momentum in that frame. The mean mediator flight distance there obeys
$\overline\ell_X^{\,*}
=
\frac{|\mathbf p_X^*|}{\mX}\ctau_X
\propto
\delta^{1/2}\ctau_X,
\qquad
\delta\to0^+.$
The laboratory flight does not vanish. At fixed parent boost and $\delta\to0^+$, the mediator inherits the heavy-state velocity: $p_X^\mu\to(\mX/\mH)p_H^\mu$, where $p_X^\mu$ and $p_H^\mu$ are the mediator and heavy-state four-momenta. Thus,
$\lim_{\delta\to0^+}\overline\ell_X
=\beta_H\gamma_H\ctau_X. $
The two flight directions also become parallel. At fixed $\ctau_{\chiH}$, the mean visible displacement then approaches $\beta_H\gamma_H(\ctau_{\chiH}+\ctau_X)$. This is a kinematic limit, not a guarantee of on-shell dominance. Figure~\ref{fig:observable_displacement} shows that the mediator flight remains important even for small $\delta$.

The two parameter choices in Fig.~\ref{fig:observable_displacement} have the same two-body proper length but different mass excesses and couplings. In the small-recoil limit at $\gamma_H=3$, the mediator still contributes $\sqrt{8}\,\ctau_X\simeq3.22$ mm to the mean displacement. Its contribution cannot be dropped.

We next calculate the electron--positron kinematics. In the $\chiH$ rest frame, the mediator energy and momentum are
\begin{equation}
E_X^*=
\frac{\mH^2-\mL^2+\mX^2}{2\mH},
\qquad
|\mathbf p_X^*|=
\frac{\lambda^{1/2}(\mH^2,\mL^2,\mX^2)}{2\mH}.
\label{eq:Xkinematics}
\end{equation}
In the $X$ rest frame, each lepton has energy $E_e^{(X)}=\mX/2$. In the $\chiH$ rest frame, the electron energy lies in
\begin{equation}
E_{e,-}^*\leq E_e^*\leq E_{e,+}^*,\qquad
E_{e,\pm}^*
=
\frac{1}{2}
\left[
E_X^*
\pm
|\mathbf p_X^*|
\sqrt{1-\frac{4m_e^2}{\mX^2}}
\right].
\end{equation}
The superscripts $*$ and $(X)$ label the $\chiH$ and $X$ rest frames, respectively. For an on-shell mediator, the pair invariant mass is $m_{ee}=\mX$.

The vector interaction also fixes the longitudinal and transverse polarization fractions of $X$~\cite{JacobWick1959}. For an unpolarized $\chiH$,
\begin{equation}
F_L=
\frac{(\mH+\mL)^2}{(\mH+\mL)^2+2\mX^2},
\qquad
F_T=
\frac{2\mX^2}{(\mH+\mL)^2+2\mX^2}.
\label{eq:polarizationfractions}
\end{equation}
Here $F_T$ sums the two transverse helicities and $F_L+F_T=1$. Let $\theta_e^{(X)}$ be the electron angle in the mediator rest frame relative to the $X$ flight direction in the $\chiH$ rest frame. Define the lepton momentum in the $X$ rest frame as $p_e^{(X)}=\frac{\mX}{2}\sqrt{1-\frac{4m_e^2}{\mX^2}}$. The normalized probability densities $f_L$ and $f_T$ in $\cos\theta_e^{(X)}$ are~\cite{Faccioli2010}
\begin{align}
f_L(\cos\theta_e^{(X)})
&=
\frac{3}{2}
\frac{m_e^2+\bigl[p_e^{(X)}\bigr]^2(1-\cos^2\theta_e^{(X)})}
{3m_e^2+2\bigl[p_e^{(X)}\bigr]^2},
\nonumber\\
f_T(\cos\theta_e^{(X)})
&=
\frac{3}{4}
\frac{2m_e^2+\bigl[p_e^{(X)}\bigr]^2(1+\cos^2\theta_e^{(X)})}
{3m_e^2+2\bigl[p_e^{(X)}\bigr]^2}.
\label{eq:angularshapes}
\end{align}

The electron energy in the $\chiH$ rest frame depends on $\theta_e^{(X)}$. The electron speed in the $X$ rest frame, in units of $c$, is $\beta_e=\sqrt{1-4m_e^2/\mX^2}$. The normalized energy spectrum is
\begin{equation}
\frac{dP}{dE_e^*}
=
\frac{2}{|\mathbf p_X^*|\beta_e}
\left[
F_L f_L(\cos\theta_e^{(X)})
+
F_T f_T(\cos\theta_e^{(X)})
\right],
\qquad
\cos\theta_e^{(X)}
=
\frac{2E_e^*-E_X^*}
{|\mathbf p_X^*|\beta_e}.
\label{eq:energyspectrum}
\end{equation}
The spectrum is normalized over $[E_{e,-}^*,E_{e,+}^*]$ and also applies to the positron. It directly connects mediator polarization to lepton energy sharing.

\begin{figure}[H]
\centering
\includegraphics[width=0.98\textwidth]{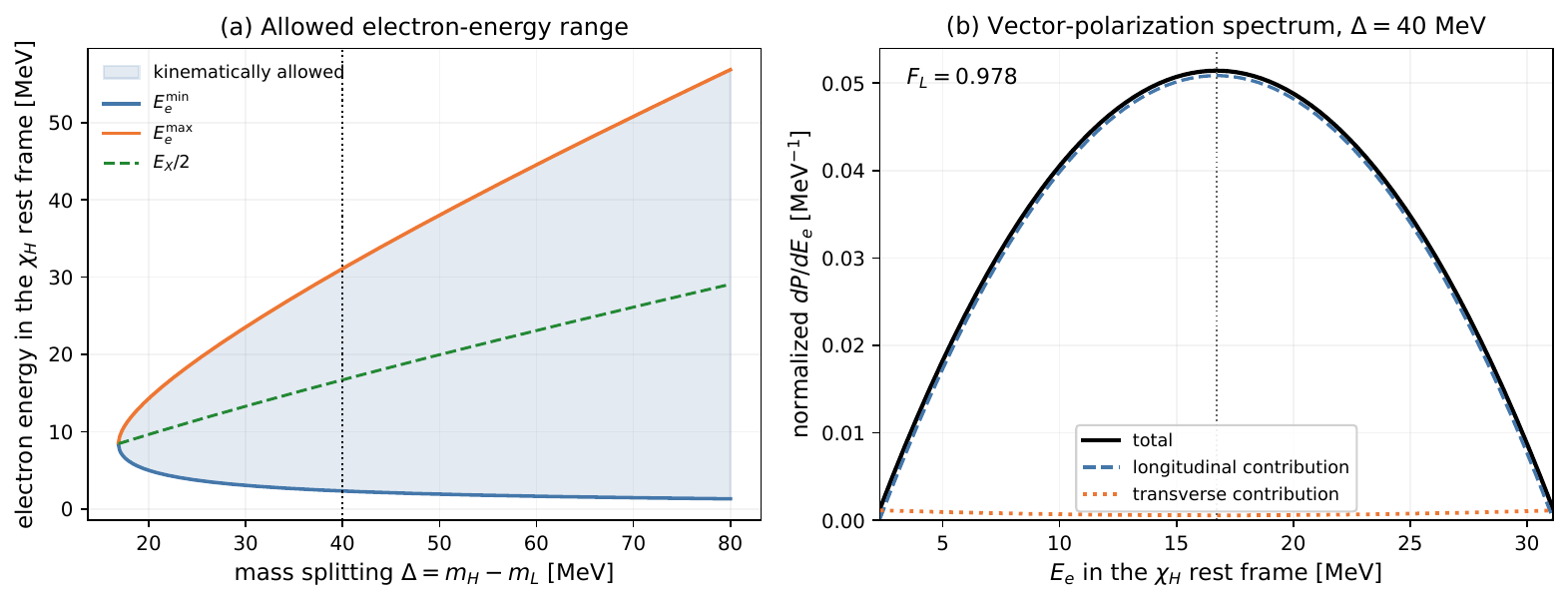}
\caption{Left: electron-energy range in the $\chiH$ rest frame versus $\DeltaM$ for $\mL=60$ MeV and $\mX=16.88$ MeV. The vertical line marks $\DeltaM=40$ MeV. Right: normalized electron-energy spectrum at this splitting, with longitudinal and transverse contributions. Here $F_L=0.978$.}
\label{fig:visiblekinematics}
\end{figure}

Figure~\ref{fig:visiblekinematics} shows these effects for $\mL=60$ MeV and $\mX=16.88$ MeV. As $\DeltaM$ approaches $\mX$ from above, the mediator recoil vanishes and both electron-energy bounds approach $\mX/2$. For $(\mH,\mL,\mX)=(100,60,16.88)$ MeV, the longitudinal fraction is $F_L=0.9782$. The energy spectrum is therefore dominated by longitudinal polarization and is not uniform.

\section{Phenomenology and Belle II geometry}
\label{sec:phenomenology}

In $e^+e^-$ collisions, an off-shell vector $X^*$ produces
$\chiH\bar\chiL$ or $\chiL\bar\chiH$. The heavy state then decays into
$\chiL X$, or its charge-conjugate final state. The on-shell mediator
subsequently decays as $X\to e^+e^-$. Both channels give the final state
$e^+e^-\to\chiL\bar\chiL e^+e^-$. The dark fermions escape detection.
The electron--positron pair identifies the single visible vertex.

We distinguish associated production from the complete resonant signal.
Their cross sections are
$\sigma_{HL}
\equiv
\sigma(e^+e^-\to\chiH\bar\chiL)
+
\sigma(e^+e^-\to\chiL\bar\chiH).$
and
$\sigma_{\rm sig}
\equiv
\sigma(e^+e^-\to\chiL\bar\chiL e^+e^-)_{\rm resonant}.$
Here $\sigma_{HL}$ includes both heavy--light production channels.
The quantity $\sigma_{\rm sig}$ includes their subsequent decays but no
detector effects. The heavy state is narrow in the displaced region:
$\Gamma_{\chiH}/\mH<3\times10^{-13}$ for a proper decay length of $1$ cm
and the masses used below. Here $\Gamma_{\chiH}$ is its total decay width.
We therefore factorize production and decay in the narrow-width approximation~\cite{UhlemannKauer2009,Fuchs2015}:
$\sigma_{\rm sig}
\simeq
\sigma_{HL}\,
{\rm Br}(\chiH\to\chiL X)\,
{\rm Br}(X\to e^+e^-),$
where ${\rm Br}$ denotes a decay branching fraction. The same expression
applies to the charge-conjugate heavy-state decay. We neglect interference
between the two resonance histories. For our reference masses, the two
heavy-state resonance conditions cannot hold simultaneously.

The interactions in Eq.~\eqref{eq:darklag} give
\begin{align}
\sigma_{HL}(s)
={}&
\frac{e^2\eps_e^2g_D^2}{12\pi s^2D_X(s)}
\lambda^{1/2}(s,\mH^2,\mL^2)
\left[s-(\mH-\mL)^2\right]
\left[2s+(\mH+\mL)^2\right],
\label{eq:productioncrosssection}
\\
D_X(s)
={}&
(s-\mX^2)^2+\mX^2\Gamma_X^2.
\nonumber
\end{align}
We neglect the electron mass in the production amplitude. The variable
$s=(p_{e^-}+p_{e^+})^2$ is the squared center-of-mass energy, with
$p_{e^-}$ and $p_{e^+}$ the incoming four-momenta. The function $D_X(s)$
contains the mediator mass $\mX$ and total width $\Gamma_X$, defined in
Sec.~\ref{sec:theory}. The K\"all\'en function $\lambda$ was defined in that
section. Equation~\eqref{eq:productioncrosssection} sums the two
charge-conjugate channels.

In the center-of-mass frame, the heavy-state energy and momentum magnitude are
\begin{equation}
E_H^{\rm cm}
=
\frac{s+\mH^2-\mL^2}{2\sqrt{s}},
\qquad
p_H^{\rm cm}
\equiv
|\mathbf p_H^{\rm cm}|
=
\frac{\lambda^{1/2}(s,\mH^2,\mL^2)}
{2\sqrt{s}}.
\label{eq:productionkinematics}
\end{equation}
The light state has momentum $\mathbf p_L^{\rm cm}=-\mathbf p_H^{\rm cm}$
and energy
$E_L^{\rm cm}
=
\frac{s+\mL^2-\mH^2}{2\sqrt{s}}.$
The superscript ${\rm cm}$ denotes the center-of-mass frame. Let
$\theta_H$ be the angle between the heavy-state momentum and the incident
electron direction in that frame. The normalized production distribution is
\begin{equation}
\frac{1}{\sigma_{HL}}\frac{d\sigma_{HL}}{d\cos\theta_H}
=
\frac{A+B\cos^2\theta_H}{2A+2B/3},
\qquad
A=E_H^{\rm cm}E_L^{\rm cm}+\mH\mL,
\qquad
B=(p_H^{\rm cm})^2,
\label{eq:productionangular}
\end{equation}
where $A$ and $B$ are coefficients specific to this angular distribution.

For Belle~II we use electron and positron beam energies of $7.004$ and
$4.002$ GeV, respectively. The half crossing angle is $41.5$ mrad relative
to the detector $z$ axis~\cite{Duerr2020}. This axis bisects the electron
beam direction and the reversed positron beam direction. These inputs give
$\sqrt{s}=10.5796$ GeV. We take an integrated luminosity
$\mathcal L=50\,\mathrm{ab}^{-1}$~\cite{BelleIIPhysicsBook2019}.
We generate production angles in the center-of-mass frame and boost the
events to the laboratory. For unpolarized beams and the pure-vector
interactions considered here, the produced $\chiH$ is unpolarized at tree
level. We verified this using spin-density traces. Its decay is therefore
isotropic in its rest frame. We retain the mediator polarization in
$X\to e^+e^-$ through Eq.~\eqref{eq:angularshapes}. We omit initial-state
radiation and beam-energy spread. The spin statement does not apply
unchanged to axial couplings or polarized beams.

The number of produced heavy--light pairs is
$N_{HL}=\mathcal L\sigma_{HL}$. The number of complete resonant cascades
before detector effects is $N_{\rm casc}=\mathcal L\sigma_{\rm sig}$.
We first impose the two-body proper length
$\ctau_{\chiH}^{(2)}=10$ mm and solve for the mass excess $\delta$ as a
function of $g_D$. At $g_D=0.1$ we find $\delta=8.5419$ eV and
$N_{HL}=35.75$--$102.39$ over the electron-coupling range defined in
Sec.~\ref{sec:theory}. We assess the correction to the physical lifetime below.

For a geometry comparison with Ref.~\cite{Duerr2020}, we place the
production point at the nominal interaction point, $\mathbf r_0=\mathbf 0$.
The beam axis defines the detector $z$ direction. Let
$R_{xy}=\sqrt{x_{ee}^2+y_{ee}^2}$ be the transverse distance of the visible
vertex from this axis. Here $x_{ee}$, $y_{ee}$ and $z$ are its
laboratory coordinates. We use two spatial regions:
$\mathcal R_{\rm low}:\quad
2~{\rm mm}<R_{xy}\leq9~{\rm mm},
 -550~{\rm mm}\leq z\leq1400~{\rm mm},$
$\mathcal R_{\rm high}:\quad
170~{\rm mm}<R_{xy}\leq600~{\rm mm},
 -550~{\rm mm}\leq z\leq1400~{\rm mm}.$
These regions exclude the intermediate radial interval. They are not the
complete selection of Ref.~\cite{Duerr2020}, which also requires a tagged
photon and a vertex-angle cut.

\begin{figure}[H]
\centering
\includegraphics[width=0.90\textwidth]{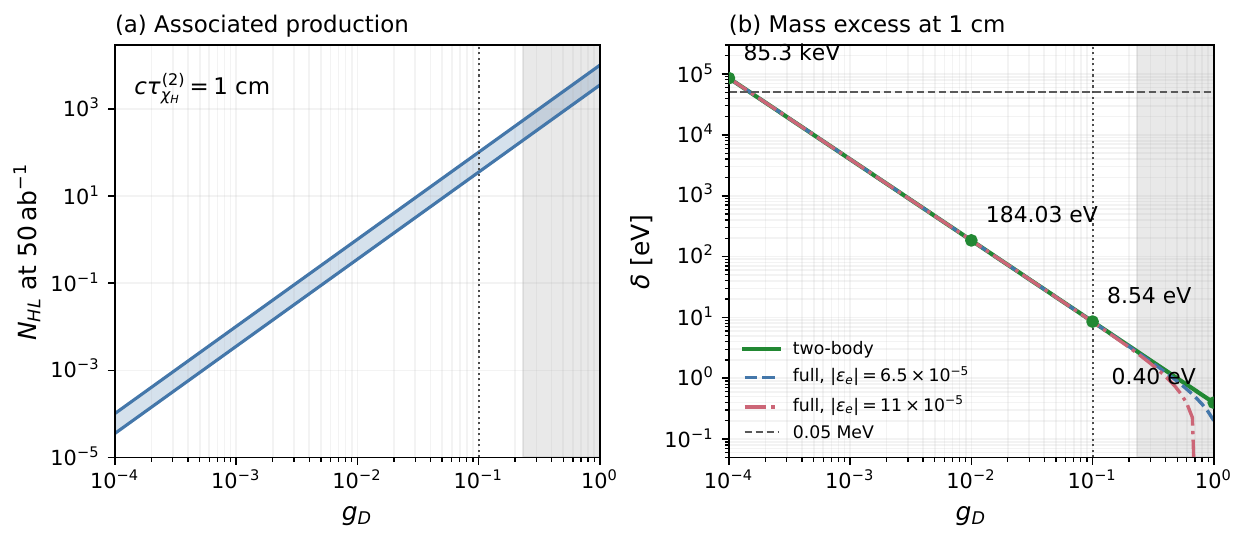}
\caption{Left: heavy--light production yield $N_{HL}$ at Belle~II for
$50\,\mathrm{ab}^{-1}$, $\mL=60$ MeV and $\mX=16.88$ MeV. The band spans
$6.5\times10^{-5}\leq|\eps_e|\leq1.1\times10^{-4}$. Right: mass excess
required for a $1$ cm proper length. The green curve uses the two-body
width. The blue and red curves use the inclusive width at the two limits of
$|\eps_e|$. The horizontal line marks the quoted mediator-mass uncertainty,
not an exclusion. Shading marks points where the off-shell correction
exceeds $10\%$ for at least one electron coupling. No detector efficiency
is applied.}
\label{fig:belle2rate}
\end{figure}

\textbf{Finite-width effects.}
A mass excess larger than the mediator width does not guarantee that the
on-shell decay dominates. Although $\delta/\Gamma_X>1.7\times10^4$ at
$g_D=0.1$, the off-shell contribution can still matter. The two-body
width is strongly suppressed at small $\delta$~\cite{UhlemannKauer2009}.
We calculate the inclusive $\chiH\to\chiL e^+e^-$ width:
\begin{equation}
\Gamma_{ee}^{\rm full}=\int_{4m_e^2}^{(\mH-\mL)^2}\!dt\,
\frac{\Gamma_2(\sqrt{t})\sqrt{t}\,\Gamma_{ee}^{X}(\sqrt{t})}
{\pi\bigl[(t-\mX^2)^2+\mX^2\Gamma_X^2\bigr]}.
\label{eq:fullwidthcheck}
\end{equation}
Here $t=m_{ee}^2$ is the squared electron--positron invariant mass.
The function $\Gamma_2(q)$ is the two-body expression in
Eq.~\eqref{eq:gammaH} evaluated at a vector mass $q=\sqrt{t}$.
Likewise, $\Gamma_{ee}^{X}(q)$ is the electron--positron width in
Eq.~\eqref{eq:gammaX} evaluated at $q$. The propagator uses the constant
on-shell width $\Gamma_X$. The integral includes the resonant contribution;
we do not add $\Gamma_2$ separately.

For $g_D=0.1$, we find
$\Gamma_{ee}^{\rm full}/\Gamma_2=1.0064$--$1.0182$ across the selected
electron couplings. The physical proper length is then
$9.82$--$9.94$ mm, rather than the nominal $10$ mm. The on-shell fraction,
approximated by $\Gamma_2/\Gamma_{ee}^{\rm full}$, is
$0.982$--$0.994$. It reduces the expected resonant cascade yield to
$35.53$--$100.56$ before detector effects. The two-body propagation
approximation is therefore accurate within about $2\%$ for this reference
configuration.

At $g_D=1$, the same two-body curve gives
$\Gamma_{ee}^{\rm full}/\Gamma_2=1.64$--$2.82$ and a physical proper
length of $3.54$--$6.11$ mm. The assumed $1$ cm lifetime is no longer
valid. The inclusive-width curves in Fig.~\ref{fig:belle2rate} instead
solve $\Gamma_{ee}^{\rm full}=\hbar c/(10~\mathrm{mm})$.
At $g_D=0.1$ they require $\delta=8.44$--$8.51$ eV, close to the
nominal $8.54$ eV used for the event sample. For
$|\eps_e|=1.1\times10^{-4}$, no positive-$\delta$ solution with a
$1$ cm physical proper length exists above $g_D=0.741$.
The corresponding limit for $|\eps_e|=6.5\times10^{-5}$ is $g_D=1.254$.
These limits follow from the nonzero off-shell width at $\delta=0$.
The inclusive integral was checked against independent numerical
integrations and the three-body amplitude.

\begin{figure}[H]
\centering
\includegraphics[width=0.66\textwidth]{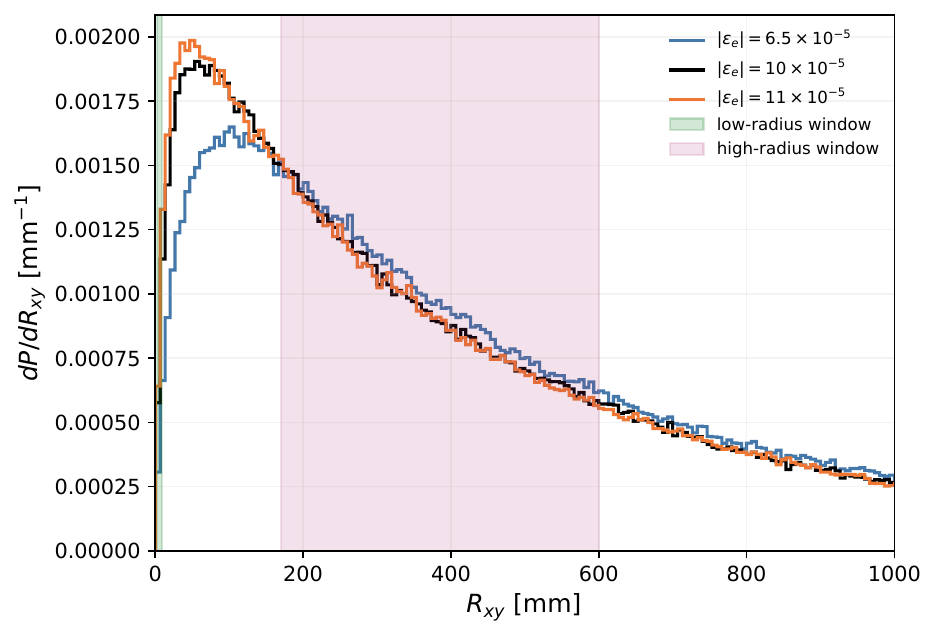}
\caption{Transverse distribution of the visible vertex at Belle~II for
$g_D=0.1$ and the nominal two-body length
$\ctau_{\chiH}^{(2)}=1$ cm. The three curves use different electron
couplings. The distribution is normalized to all
generated decays, including those outside the plotted range. No trigger,
reconstruction or detector efficiency is included.}
\label{fig:belle2lab}
\end{figure}

Figure~\ref{fig:belle2lab} shows how the electron coupling affects the
visible-vertex position. Smaller $|\eps_e|$ gives a longer mediator
lifetime and shifts the vertices outward. Both neutral flights contribute.
The radial distribution alone cannot determine $g_D$ and $|\eps_e|$
separately.

\subsection{Laboratory lepton kinematics}
\label{sec:lableptons}

We generate $N=3\times10^6$ unweighted on-shell cascades for the
reference choice $g_D=0.1$ and $\ctau_{\chiH}^{(2)}=10$ mm.
The masses are $\mH=76.88000854$ MeV, $\mL=60$ MeV and
$\mX=16.88$ MeV. The random seed is 20261012. We include both
charge-conjugate production channels and the electron mass in all decays
and boosts. The pair energy in the laboratory is
$E_{ee}=E_{e^-}^{\rm lab}+E_{e^+}^{\rm lab}$. We denote its invariant
mass by $m_{ee}$. The opening angle $\theta_{ee}$ is the angle between
the laboratory lepton momenta. We define the softer-lepton momentum as
$p_{\rm soft}=\min(|\mathbf p_{e^-}^{\rm lab}|,|\mathbf p_{e^+}^{\rm lab}|)$.
The superscript ${\rm lab}$ denotes the laboratory frame.

The mediator has only $|\mathbf p_X^*|=0.0150$ MeV of momentum in the
heavy-state rest frame. It therefore inherits most of the heavy-state
boost. We obtain $0.87446\leq E_{ee}/{\rm GeV}\leq1.54270$ and
$51.80\leq\gamma_X\leq91.39$, where $\gamma_X=E_{ee}/\mX$ is the
mediator Lorentz factor. The electron--positron pair is strongly collimated.
Large opening angles mainly occur when one lepton has low momentum.

The displaced-pair proposal of Ref.~\cite{Duerr2020} requires both
leptons to have momenta above $0.1$ GeV, an opening angle above $0.1$ rad,
and an invariant mass above $30$ MeV. That proposal also requires a
photon. Our Born-level signal contains no such photon. A published Belle~II
dark-Higgs search~\cite{BelleII2025IDM} uses a similar angle cut but a
different final state. Neither analysis sets a universal limit on the
reconstruction of collimated pairs at Belle~II.

For our reference configuration, the individual passing fractions are
$P(p_{\rm soft}>0.1~\mathrm{GeV})=0.83059\pm0.00022,
\qquad
P(\theta_{ee}>0.1)=0.03999\pm0.00011.$
Here $P$ denotes the fraction of generated events satisfying a cut.
The quoted uncertainties are binomial standard errors
$\sqrt{\hat p(1-\hat p)/N}$, with $\hat p$ the measured fraction.
No simulated event satisfies both cuts. The invariant-mass requirement
also rejects every on-shell event: $m_{ee}=\mX=16.88$ MeV.

The zero joint acceptance is not a Monte Carlo artifact. It follows from
energy--momentum conservation. Let $p_s=p_{\rm soft}$ and $p_h$ be the
softer- and harder-lepton momentum magnitudes. Their energies are $E_s$
and $E_h$. Then
\begin{equation}
\mX^2=2m_e^2+2(E_sE_h-p_sp_h\cos\theta_{ee})
\geq4m_e^2+2p_sp_h(1-\cos\theta_{ee}),
\label{eq:massanglebound}
\end{equation}
since $E_sE_h-p_sp_h\geq m_e^2$.
Let $p_0>0$ and $\theta_0>0$ be the minimum accepted lepton
momentum and opening angle. Define
$K=(\mX^2-4m_e^2)/[2(1-\cos\theta_0)]$.
The cuts $p_s\geq p_0$ and $\theta_{ee}\geq\theta_0$ imply
$p_sp_h\leq K$. If $K<p_0^2$, no event can pass. Otherwise the largest pair energy
occurs at $p_s=p_0$ and $p_h=K/p_0$. Hence
\begin{equation}
E_{ee}\leq\sqrt{p_0^2+m_e^2}+\sqrt{(K/p_0)^2+m_e^2}
\equiv E_{\rm cap}(p_0,\theta_0).
\label{eq:energycutbound}
\end{equation}
Here $E_{\rm cap}$ is the maximum pair energy compatible with both
requirements. At $p_0=0.1$ GeV and $\theta_0=0.1$ rad,
$E_{\rm cap}=0.38413$ GeV. The signal has
$E_{ee}\geq E_{ee}^{\min}=0.87446$ GeV, from the production
kinematics and successive boosts specified above. The two cuts are therefore
incompatible for this on-shell, Born-level configuration, even without
the mass cut. The conclusion need not hold for different masses,
radiative production, off-shell decays or detector smearing.

We test other momentum and angular requirements through the joint
kinematic acceptance
$A_{\rm kin}(p_0,\theta_0)=P(p_{\rm soft}>p_0,\ \theta_{ee}>\theta_0).$
This is a particle-level fraction. It does not include geometry,
trigger or reconstruction.

\begin{figure}[t]
\centering
\includegraphics[width=0.98\textwidth]{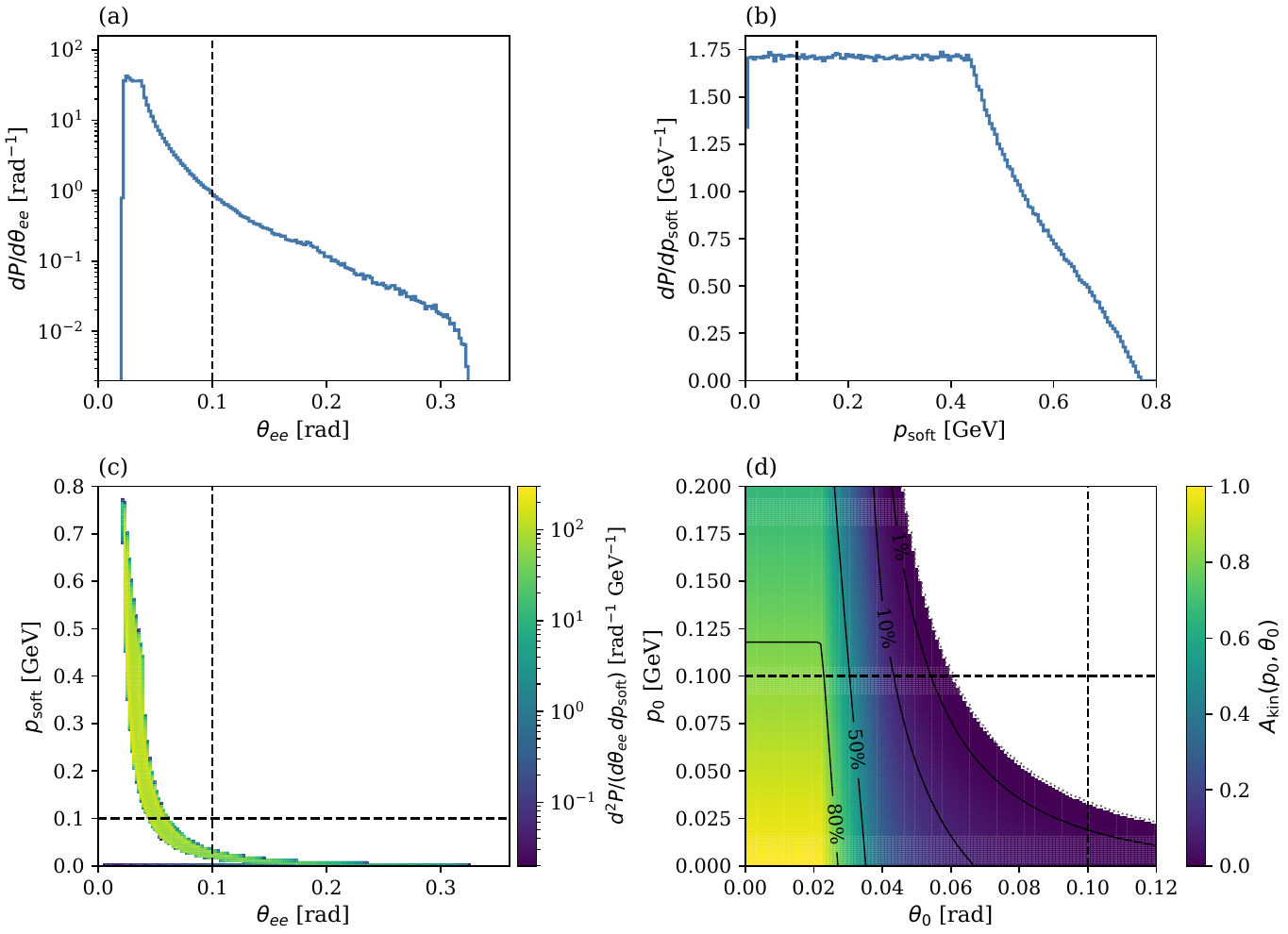}
\caption{Laboratory lepton kinematics for the reference mass choice.
Panels (a--c) show the normalized opening-angle and softer-lepton
momentum distributions, and their joint density, from
$3\times10^6$ generated on-shell cascades. Panel (d) shows
$A_{\rm kin}(p_0,\theta_0)$ without a pair-mass requirement or detector
effects. Solid curves mark $1$, $10$, $50$ and $80\%$ acceptance. White
cells have no accepted events. The dotted boundary follows from
Eq.~\eqref{eq:energycutbound} or $K<p_0^2$; dashed lines mark the
reference momentum and angular cuts.}
\label{fig:lableptons}
\end{figure}

Figure~\ref{fig:lableptons}(a--c) shows why the lepton cuts compete:
large opening angles mainly select events with a soft lepton.
Figure~\ref{fig:lableptons}(d) scans
$0\leq p_0\leq0.20$ GeV and $0\leq\theta_0\leq0.12$ rad in increments
of $0.001$ GeV and $0.001$ rad. If $p_0=0$ or $\theta_0=0$,
the corresponding cut is absent; the analytic expression for
$E_{\rm cap}$ applies when both cuts are positive.
No invariant-mass cut is imposed.
At $p_0=0.1$ GeV, an angle requirement of $\theta_0=0.03$ rad gives
$A_{\rm kin}=(51.602\pm0.029)\%$; increasing it to $0.05$ rad reduces
the acceptance to $(2.965\pm0.010)\%$. At $\theta_0=0.1$ rad, even
lowering the momentum requirement to $p_0=0.02$ GeV gives only
$(0.850\pm0.005)\%$ acceptance. These values do not establish
experimentally viable selections.

The region with $E_{\rm cap}<E_{ee}^{\min}$ or $K<p_0^2$ has exactly
zero kinematic acceptance. The remaining map is based on the generated
events. Its pointwise statistical error is at most $2.89\times10^{-4}$.
An independent zero-recoil integration agrees with the Monte Carlo
result: at $(p_0,\theta_0)=(0.1~\mathrm{GeV},0.03~\mathrm{rad})$ it
gives $0.51583$ for the acceptance. Additional checks reproduce the
production cross section and polarization distributions.

At $|\eps_e|=10^{-4}$, the nominal propagation model places $0.569\%$
and $37.369\%$ of visible vertices in $\mathcal R_{\rm low}$ and
$\mathcal R_{\rm high}$, respectively. These are geometric fractions,
not detection efficiencies. A dedicated search needs a trigger and track
reconstruction suitable for low-mass, collimated $e^+e^-$ pairs~\cite{Lees2015LongLived,Ferber2022,Duerr2021}.
Material interactions, photon conversions, track separation and magnetic
transport must be studied with detector simulation. We do not infer a
detector-level sensitivity.

\section{Conclusions}
\label{sec:conclusions}

We investigated displaced $e^+e^-$ signatures from an inelastic dark sector mediated by a light vector boson in the $X_{17}$ mass region. We calculated the decay widths, the distribution of the visible vertex, and the polarization-dependent lepton kinematics. We also determined the associated production rate in electron--positron collisions and studied the resulting signatures at Belle~II.

Our results reveal a tension between production rate and decay length. Increasing the dark-sector coupling enhances production but shortens the heavy-state lifetime at fixed masses. Maintaining a macroscopic displacement requires an increasingly small mass excess. This relation reaches the eV scale and has no established symmetry protection in the effective model. We also quantified off-shell contributions and identified where the two-body approximation fails. The visible displacement depends on two consecutive neutral flight paths, even when the mediator recoil becomes negligible.

At Belle~II, we demonstrated analytically that the reference momentum and opening-angle cuts are incompatible with the signal for the mass configuration studied. Alternative selections recover substantial kinematic acceptance. In particular, reducing the opening-angle requirement allows approximately half of the generated events to pass the combined momentum and angular cuts, without an invariant-mass cut. These results identify the kinematic conditions needed for a dedicated low-mass search. Detector reconstruction, backgrounds, and trigger requirements must be evaluated before estimating experimental sensitivity.

\section*{Author Contributions}
Conceptualization, R.S. and A.C.; methodology and formal analysis, R.S. and A.C.; validation, R.S. and A.C.; writing---original draft preparation, R.S. and A.C.; writing---review and editing, R.S. and A.C. All authors have read and agreed to the submitted version of the manuscript.

\section*{Acknowledgments}
R.S. acknowledges support from the project ``Implementation of cutting-edge research and its application as part of the Scientific Center of Excellence for Quantum and Complex Systems, and Representations of Lie Algebras'', Grant No. PK.1.1.10.0004, co-financed by the European Union through the European Regional Development Fund - Competitiveness and Cohesion Programme 2021-2027.
A.C. acknowledges partial support from MIUR and INFN. A.C. also acknowledges COST Action CA1511, Cosmology and Astrophysics Network for
Theoretical Advances and Training Actions (CANTATA).

\section*{Data Availability Statement}
No experimental datasets were used.

\section*{Conflicts of Interest}
The authors declare no conflict of interest.

\sloppy
\bibliographystyle{unsrt_boldvolume}
\bibliography{references}

\end{document}